\documentclass[english,prl,twocolumn,showpacs,amsmath,amssymb,superscriptaddress,floatfix]{revtex4}
\usepackage[T1]{fontenc}
\usepackage[latin1]{inputenc}
\usepackage{graphicx}
\makeatletter

\usepackage{bm}
\usepackage{pifont}

\usepackage{babel}
\makeatother

\begin{document}


\title{Transport Spectroscopy of a Graphene Quantum Dot Fabricated by Atomic Force Microscope Nanolithography}

\author{R.K. Puddy}
\affiliation{Cavendish Laboratory, University of Cambridge, Cambridge CB3 0HE, United Kingdom}
\author{C.J. Chua}
\affiliation{Cavendish Laboratory, University of Cambridge, Cambridge CB3 0HE, United Kingdom}
\author{M.R. Buitelaar}\email{m.buitelaar@ucl.ac.uk}
\affiliation{Cavendish Laboratory, University of Cambridge, Cambridge CB3 0HE, United Kingdom}
\affiliation{London Centre for Nanotechnology, UCL, London WC1H 0AH, United Kingdom}
\affiliation{Department of Physics and Astronomy, UCL, London WC1E 6BT, United Kingdom}


\begin{abstract}
We report low-temperature transport spectroscopy of a graphene quantum dot fabricated by atomic force microscope nanolithography. The excellent spatial resolution of the atomic force microscope allows us to reliably fabricate quantum dots with short constrictions of less than 15 nm in length. Transport measurements demonstrate that the device is dominated by a single quantum dot over a wide gate range. The electron spin system of the quantum dot is investigated by applying an in-plane magnetic field. The results are consistent with a Land\'{e} $g$-factor $\sim 2$ but no regular spin filling sequence is observed, most likely due to disorder.
\end{abstract}




\maketitle

The electronic properties of graphene are of intense current interest because of unique features such as the two-dimensional
nature of the graphene lattice and its gapless Dirac spectrum. These properties give rise to novel phenomena such as an anomalous half-integer quantum Hall effect \cite{Novoselov,Zhang1}, Klein tunneling \cite{Katsnelson}, and an optical transmittance defined solely by the fine-structure constant \cite{Nair}. Another attraction is the absence of nuclear spin in the dominant carbon-12 isotope, offering the possibility to define spin qubits in graphene quantum dots which do not suffer from the decoherence associated with the hyperfine interaction \cite{Trauzettel}. The spin system of graphene quantum dots however, is not yet properly understood \cite{Ponomarenko,Guttinger}.

The absence of a band gap in the electronic spectrum and the existence of Klein tunneling means that devices such as quantum dots cannot be defined electrostatically in monolayer graphene. It is possible to introduce a gap in the spectrum of bilayer graphene by applying an electric field and thus define quantum dots using gate electrodes. Typical energies of quantum dots defined in bilayer graphene, however, are too small to readily observe the quantization of the electronic spectrum \cite{Allen,Goossens}. A confinement gap can be introduced in monolayer graphene by physically defining narrow constrictions in the material. These constrictions can be used as tunable barriers to graphene quantum dots \cite{Han}. It has been observed however, that in long constrictions, a confinement gap in combination with a background disorder potential produces a hard transport gap where conduction occurs via of a number of quantum dots in series in the barriers \cite{Stampfer,Todd}. This complicates the transport characteristics and prevents strong coupling between the defined quantum dots  and the graphene source and drain electrodes. It is therefore of considerable interest to be able to fabricate graphene quantum dots with short constrictions \cite{Terres}.

In this work we explore a fabrication technique - atomic force microscope (AFM) nanolithography - which, due to its excellent spatial resolution allows us to reliably fabricate very short constrictions ($L \sim 15$ nm) between the graphene quantum dots and electrodes. An additional advantage of this fabrication method is that the graphene is not exposed to photoresists and therefore remains clean. Here we present low-temperature magneto-transport spectroscopy of a graphene quantum dot fabricated by AFM nanolithography. From our measurements we find the electron addition spectrum and investigate the spin system in an in-plane magnetic field.

\begin{figure}
\includegraphics[width=80mm]{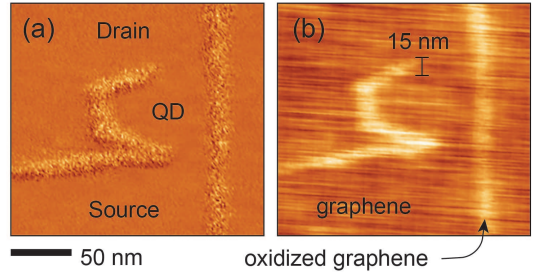}
\caption{\label{Fig1} (color online)  \textbf{(a)} Atomic force microscope (AFM) friction image and \textbf{(b)} AFM height image of a single quantum dot fabricated in monolayer graphene using AFM nanolithography. The spatial resolution of the technique is $\sim$ 15 nm for the oxidized lines. The bumps are about 1.5 nm in height.}
\end{figure}

The device we consider, shown in Fig. 1, consists of a quantum dot of $\sim 70$ nm diameter defined by oxidized lines on a monolayer graphene sheet. The graphene, supported on a doped Si wafer with a 300 nm SiO$_2$ capping layer, was identified as monolayer using optical contrast \cite{contrast}. The oxidized lines were created using an atomic force microscope with a negative voltage of between -5 V and -7 V applied to the AFM tip with the sample grounded \cite{Weng,Masubuchi,Neubeck,Byun}. We used doped Si contact mode tips in ambient conditions with a relative humidity of approximately 40-50 $\%$. Forces were minimized during lithography, typically less than 5 nN \cite{manufacturer}. When much larger forces were applied, using tapping mode tips, the graphene could be etched away completely \cite{Puddy}. The reliability of AFM nanolithography was excellent provided that the graphene surface had not previously been exposed to photoresists. For this reason we used shadow masking to electrically contact the graphene prior to AFM nanolithography \cite{Bao}. Using this method we fabricated and measured five graphene quantum dot devices of which the one studied in most detail is presented here.

\begin{figure}
\includegraphics[width=84mm]{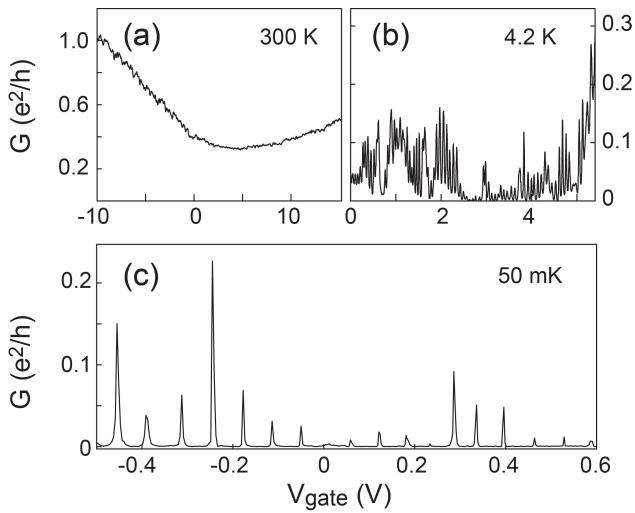}
\caption{\label{Fig2} (color online) \textbf{(a)} Conductance as a function of back gate voltage of a the graphene device measured at room temperature. \textbf{(b)} Conductance measured at $T=4.2$ K. A transport gap appears, centered around $V_{g} = 3$ V. \textbf{(c)} Conductance measured at $T \sim 50$ mK inside the transport gap showing equidistantly spaced peaks as expected for a single quantum dot dominated by Coulomb blockade.}
\end{figure}

\begin{figure}[b]
\includegraphics[width=85mm]{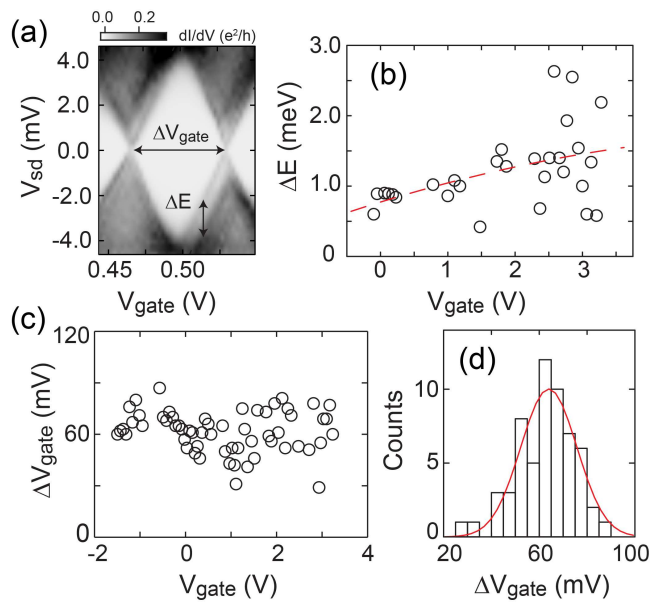}
\caption{\label{Fig3}(color online) \textbf{(a)} A representative Coulomb blockade diamond measured at $T \sim 50$ mK. From the slopes of the diamonds the various capacitances can be extracted. Excited states, with spacing $\Delta E$, are visible as lines parallel to the diamond edges, indicated by the arrow. \textbf{(b)} Excited state spacing, $\Delta E$, as a function of gate voltage. The dashed line is a guide to the eye. \textbf{(c)} Gate spacing $\Delta V_g$ between Coulomb blockade peaks, as indicated in panel (a), as a function of gate voltage. \textbf{(d)} Histogram of the data shown in panel (c) with a Gaussian fit.}
\end{figure}

\begin{figure*}
\includegraphics[width=175mm]{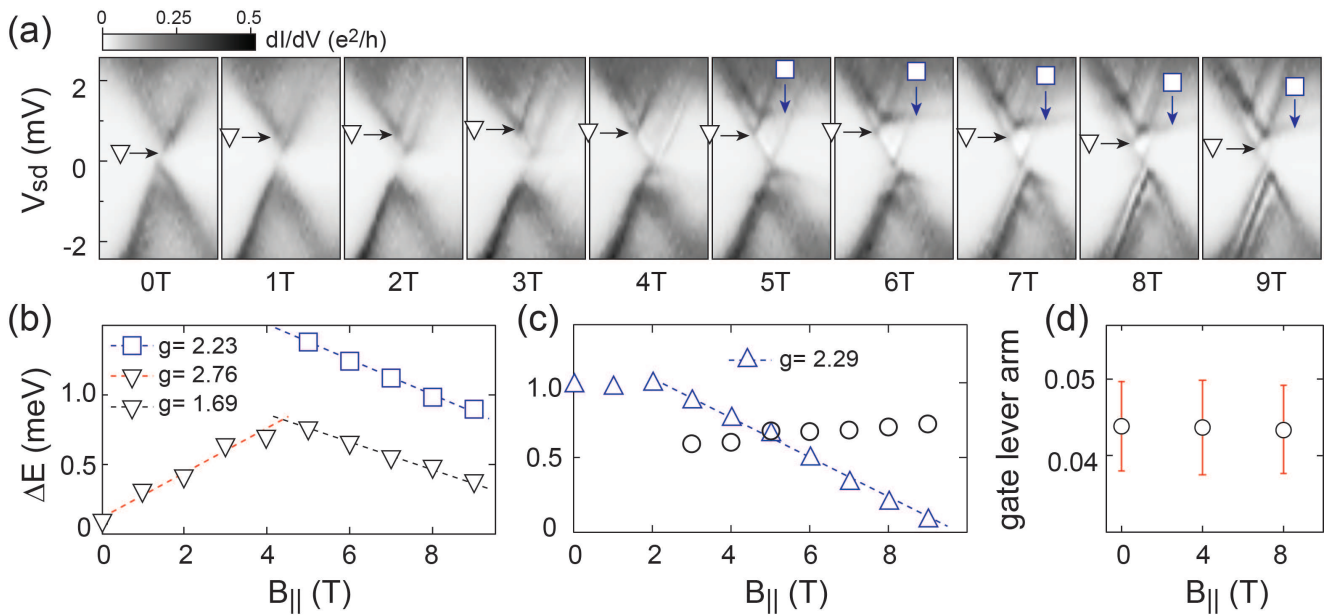}
\caption{\label{Fig4} (color online) \textbf{(a)} A Coulomb diamond measured at magnetic fields $B_\|$ from 0 to 9 T.
Excited states appear and move with magnetic field as indicated for two different excitations (triangles and squares). \textbf{(b)} Magnetic field evolution of the excitations highlighted in panel (a). \textbf{(c)} Magnetic field evolution of the excitations of a different Coulomb diamond (not shown). The dashed lines in (b) and (c) are best fit slopes and give the Land$\acute{e}$ $g$ factor. Some excitations do not move in field. \textbf{(d)} Lever arms extracted at 0, 4, and 8 Tesla showing little variation.}
\end{figure*}

Conductance as a function of back gate voltage, $V_g$, at temperatures of $T \sim 300$ K and $4.2$ K are shown in Figs. 2(a) and (b) respectively. At $T \sim 300$ K the conductance has a minimum at $V_g \sim 3$ V. At $T \sim 4.2$ K, Coulomb blockade resonances are observed on top of the conductance background and a transport gap appears which extends over several volts. The transport gap in gate voltage can be compared to the expected confinement gap at the barriers which we estimate as $E_{gap} \sim 30$ meV for our device where the barrier width $W \sim 40$ nm \cite{White}. Given a gate lever arm $\alpha_g \sim 0.05$ (obtained below), this would correspond to a transport gap over a gate voltage range of about 0.6 V. This is much less than observed in the experiment. The large transport gap in gate voltage is consistent with results on graphene nanoconstrictions of lengths down to 50 nm and has been attributed to disorder induced localized states \cite{Terres}. This work shows that these observations hold even for the 15 nm constrictions used in our work which is perhaps surprising as this is less than the typical disorder potential length scale of graphene on SiO$_2$ \cite{Martin,Zhang2}.

Figure 2(c) shows the conductance as a function of $V_g$ at $T \sim 50$ mK  which is the base temperature of our dilution refrigerator. We observe a series of equidistantly spaced Coulomb blockade peaks indicating that the transport characteristics are dominated by a single quantum dot. In Fig. 3(a) we show the differential conductance as a function of source drain voltage, $V_{sd}$, and $V_g$. We obtain the characteristic Coulomb diamond structures and observe excited states as lines parallel to the diamond slopes. We observe co-tunneling across the full gate range and note that it is weakest around $V_g \sim 3$ V. From the slopes of the diamonds we calculate the gate lever arm $\alpha_g = C_g/C_\Sigma \sim 0.05$, where $C_g$ and $C_\Sigma$ are the gate and total capacitance of the quantum dot, respectively. The lever arm increases slightly to $\sim 0.08$ with more positive $V_g$ as the source and drain capacitances to the quantum dot become smaller as the barriers become more opaque. From the spacing in gate voltage, $\Delta V_g$, between Coulomb blockade peaks, we extract the gate capacitance $C_g \sim e / \Delta V_g$ yielding $C_g \sim 2.5$ aF which is consistent with the value expected from the geometry of the device \cite{geometry}.

In Fig. 3(c) we plot $\Delta V_g$ as a function of $V_g$ and find that the average value of $\Delta V_g \sim 62$ mV, does not change appreciably as the backgate voltage is varied, indicating that $C_g$, and thus the size of the quantum dot, remains unchanged. The histogram of $\Delta V_g$, shown in Fig. 3(d), fits well with a Gaussian distribution for which we find a standard deviation of $\sim 12$ mV, corresponding to a normalised standard deviation of 0.20, in good agreement with recent findings \cite{Engels}. This variation could be due to fluctuations in the single-particle level spacing $\Delta E$, which can be found from the spacing of the first excited states, as indicated in Fig. 3(a). This is plotted for a wide gate range in Fig. 3(b). We find $\Delta E \sim 1$ meV, although with considerable scatter. Given $\alpha_g \sim 0.05$ this corresponds to $\Delta V_g \sim 20$ mV. The measured single-particle level spacing and its fluctuations are thus sufficiently large to explain the observed scatter in $\Delta V_g$. We note that, the single-particle level spacing in monolayer graphene quantum dots is theoretically expected \cite{Schnez} to show a $\Delta E \propto 1/\sqrt{N}$ dependence, where $N$ is the number of charge carriers. While the average $\Delta E$ does indeed appear to increase somewhat with decreasing $N$, the scatter in the data is too large to verify this relation accurately.

\begin{figure*}[t]
\includegraphics[width=175mm]{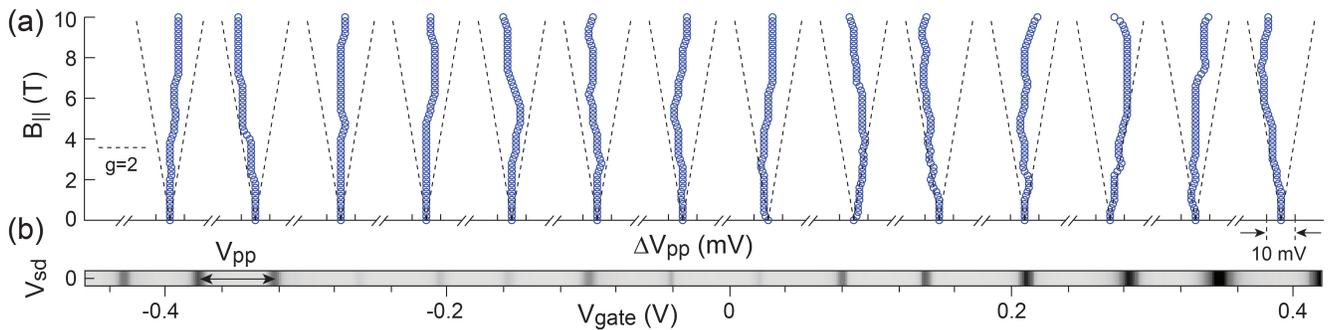}
\caption{\label{Fig5} (color online) \textbf{(a)} The change in separation between the zero-bias Coulomb blockade peaks shown in panel (b), as a function of $B_\|$. The black dashed lines represent the Zeeman energy $E_Z = g \mu_B B$ assuming $g = 2$ and a lever arm $\alpha_g = 0.05$. \textbf{(b)} Zero-bias Coulomb blockade peaks at $B_\|=0$ T.}
\end{figure*}

Finally, we turn to the spin filling sequence in our quantum dot which we investigated by applying a magnetic field of up to 10 Tesla. The field was carefully aligned to be parallel (within $\sim$ 1-2 degrees) with the graphene lattice such that the dominant contribution is the Zeeman splitting $E_Z$ of the single-particle levels. In the simplest case one might expect a simple up-down spin filling sequence although previous work suggested that the spin filling could be dominated by exchange interaction \cite{Guttinger, Guttinger2}.

To determine the spin filling in our device we first investigated the magnetic field evolution of the excited states observed in the Coulomb diamonds. Figure 4 (a) shows a Coulomb diamond, at $V_g \sim 0.3$ V, in parallel magnetic fields of $B_\| = 0 - 9$ T. Two peaks in the differential conductance, highlighted by the squares and triangles, are seen to move in field. This movement is plotted in Fig. 4(b) and for two peaks from another region of $V_g$ in Fig 4(c). The extracted slopes $\Delta E_Z/\Delta B_\|$ are consistent with a Land$\acute{e}$ $g$-factor $\sim 2$ although with significant scatter between different slopes. We verified that the lever arms remain largely unaffected by the magnetic field, see Fig. 4(d).

We then followed the spin states by measuring the Coulomb peak spacing $\Delta V_g$ at $V_{sd}=0$ V, i.e. the linear-response regime, as a function of parallel magnetic field, see Fig.5. Assuming a simple $\uparrow \downarrow \uparrow \downarrow$ spin filling sequence one expects peaks to alternately become closer and further apart with slopes of $\pm g \mu_B$. Alternatively, for a sequence $\downarrow \uparrow \uparrow \downarrow \downarrow \uparrow \uparrow \downarrow$ due to, e.g. exchange interactions, slopes of zero are expected for every second pair \cite{Guttinger}. As apparent in the data shown in Fig. 5, where $\pm g \mu_B$ is indicated by the dashed lines for $g=2$, we occasionally see slopes consistent with $0,\pm 2 \mu_B$, but no clear sequence emerges which would allow us to draw any definite conclusions. The absence of a clear spin filling sequence is similar to findings on plasma etched graphene quantum dots and is most likely due to edge and/or substrate induced disorder \cite{Kolbl}. In this respect, it would be of considerable interest to be able to suspend the graphene dots to eliminate substrate interaction. We believe this to be feasible for quantum dots fabricated by our AFM nanolithography technique as suspended dots would still have mechanical support from the surrounding graphene lattice even when electrically isolated. Suspended graphene quantum dots would also be of interest to investigate the interplay between vibrational modes and the single-electron tunneling current for which the precise spin state is less important \cite{Steele}.

In conclusion, we have used AFM nanolithography to fabricate quantum dots in monolayer graphene. The spatial resolution of the AFM tip allowed us to fabricate constrictions to the quantum dots with lengths $\sim 15$ nm. From low-temperature bias spectroscopy we investigated the addition energy and spin filling sequence in a magnetic field parallel to the graphene lattice. The measurements demonstrate that transport is dominated by a single quantum dot in this device. Measurements of the electron spin states are consistent with a Land\'{e} $g$-factor $\sim 2$ but no clear spin filling sequence was observed, most likely due to disorder. Future work will address the role of substrate interaction by suspending the graphene quantum dots.

The authors thank Malcolm Connolly and Charles Smith for helpful discussions.
This work was supported by EPSRC and the Royal Society (M.R.B.).


\begin{thebibliography}{10}

\bibitem{Novoselov}
K.S. Novoselov, A.K. Geim, S.V. Morozov, D. Jiang, M.I. Katsnelson, I.V. Grigorieva, S.V. Dubonos, and A.A. Firsov, Nature \textbf{438}, 197 (2005).

\bibitem{Zhang1}
Y. Zhang, J.W. Tan, H.L. Stormer, and P. Kim, Nature \textbf{438}, 201 (2005).

\bibitem{Katsnelson}
M.I. Katsnelson, K.S. Novoselov, and A.K. Geim, Nature Phys. \textbf{2}, 620 (2006).

\bibitem{Nair}
R.R. Nair, P. Blake, A.N. Grigorenko, K.S. Novoselov, T.J. Booth, T. Stauber, N.M.R. Peres, and A.K. Geim, Science \textbf{320}, 5881 (2008).

\bibitem{Trauzettel}
B. Trauzettel, D. Bulaev, D. Loss, and G. Burkard, Nature Phys. \textbf{3}, 192 (2007).

\bibitem{Ponomarenko}
L.A. Ponomarenko, F. Schedin, M.I. Katsnelson, R. Yang, E.W. Hill, K.S. Novoselov, and A.K. Geim, Science \textbf{320}, 356 (2008).

\bibitem{Guttinger}
J. G\"{u}ttinger, T. Frey, C. Stampfer, T. Ihn, and K. Ensslin, Phys. Rev. Lett. \textbf{105}, 116801 (2010).

\bibitem{Allen}
M.T. Allen, J. Martin, and A. Yacoby, Nat. Commun. \textbf{3}, 934 (2012).

\bibitem{Goossens}
A.M. Goossens, S.C.M. Driessen,T.A. Baart, K. Watanabe,T. Taniguchi, and L.M.K. Vandersypen, Nano Lett. \textbf{12}, 4656 (2012).

\bibitem{Han}
M.Y. Han, B. \"{O}zyilmaz, Y. Zhang, and P. Kim, Phys. Rev. Lett. \textbf{98}, 206805 (2007).

\bibitem{Stampfer}
C. Stampfer, J. G\"{u}ttinger, S. Hellm\"{u}ller, F. Molitor, K. Ensslin, and T. Ihn, Phys. Rev. Lett. \textbf{102}, 056403 (2009).

\bibitem{Todd}
K. Todd, H.-T. Chou, S. Amasha, D. Goldhaber-Gordon, Nano Lett. \textbf{9}, 416 (2009).

\bibitem{Terres}
B. Terres, J. Dauber, C. Volk, S. Trellenkamp, U. Wichmann, and C. Stampfer, Appl. Phys. Lett. \textbf{98}, 032109 (2011).

\bibitem{contrast}
We have verified the reliability of the technique by Raman spectroscopy. In all cases (>20) the optical contrast method
correctly identified mono and bi-layer graphene.

\bibitem{Weng}
L. Weng, L. Zhang, Y. P. Chen, and L. P. Rokhinson, Appl. Phys. Lett. \textbf{93}, 093107 (2008).

\bibitem{Masubuchi}
S. Masubuchi, M. Ono, K. Yoshida, and T. Machida, Appl. Phys. Lett. \textbf{94}, 082107 (2009).

\bibitem{Neubeck}
S. Neubeck, L. A. Ponomarenko, F. Freitag, A. J. M. Giesbers, U. Zeitler, S. V. Morozov, P. Blake, A. K. Geim, and K. S. Novoselov, Small \textbf{6}, 1469 (2010).

\bibitem{Byun}
I. Byun, D. Yoon, J. Choi, I. Hwang, D. Lee, M. Lee, T. Kawai, Y. Son, Q. Jia, H. Cheong, and B. Park, ACS Nano \textbf{5}, 6417 (2011).

\bibitem{manufacturer}
We used highly doped Si contact mode tips with a force constant of $\sim 0.2$ N/m.

\bibitem{Puddy}
R.K. Puddy, P.H. Scard, D. Tyndall, M.R. Connolly, C.G. Smith, G.A.C. Jones, A. Lombardo, A.C. Ferrari, and M.R. Buitelaar, Appl. Phys. Lett. \textbf{98}, 133120 (2011).

\bibitem{Bao}
W. Bao, G. Liu, Z. Zhao, H. Zhang, D. Yan, A. Deshpande, B.J. LeRoy, and C.N. Lau, Nano Research \textbf{3}, 98 (2010).

\bibitem{White}
Here we use  $\Delta E_{gap} \sim 1.2/W$ eV, where $W$ is the barrier width in nm, see e.g. C.T. White, J. Li, D. Gunlycke, and J.W. Mintmire, Nano. Lett.\textbf{7}, 825 (2007).

\bibitem{Martin}
J. Martin, N. Akerman, G. Ulbricht, T. Lohmann, J.H. Smet, K. von Klitzing, and A. Yacoby, Nature Phys. \textbf{4}, 144 (2008).

\bibitem{Zhang2}
Y. Zhang, V.W. Brar, C. Girit, A. Zettl, and M.F. Crommie, Nature Phys. \textbf{5}, 722 (2009).

\bibitem{geometry}
The quantum dot has a diameter $d \sim 70$ nm as estimated from the AFM images. The SiO$_2$ on the substrate has a $t=300$ nm thickness and dielectric constant $\epsilon_r = 3.8$. To estimate the gate capacitance we use $C_g = 2 \epsilon_0 (\epsilon_r + 1) d$ which is a good approximation for on isolated disk above a parallel plate given $d \ll t$. This yields $C_g = 5.9$ aF. The gate capacitance of a disk in plane with a good conductor is approximated by $C_g = A\epsilon_0\epsilon_r/t$ with $A$ the dot area. This yields $C_g = 0.43$ aF. The precise value for our quantum dot is expected to be within this range due to (partial) screening of the surrounding graphene, in good agreement with the measured 2.5 aF.

\bibitem{Engels}
S. Engels, A. Epping, C. Volk, S. Korte, B. Voigtl\"{a}nder, K. Watanabe, T. Taniguchi, S. Trellenkamp, and C. Stampfer, Appl. Phys. Lett. \textbf{103}, 073113 (2013).

\bibitem{Schnez}
S. Schnez, F. Molitor, C. Stampfer, J. G\"{u}ttinger, I. Shorubalko. T. Ihn, and K. Ensslin, Appl. Phys. Lett. \textbf{94}, 012107 (2009).

\bibitem{Guttinger2}
J. G\"{u}ttinger, F. Molitor, C. Stampfer, S. Schnez, A. Jacobsen, S. Dr\"{o}scher, T. Ihn, and K. Ensslin, Rep. Prog. Phys. \textbf{75}, 126502 (2012).

\bibitem{Kolbl}
D. K\"{o}lbl and D.M. Zumb\"{u}hl, arXiv:1307.8163.

\bibitem{Steele}
A. Steele, A.K. H\"{u}ttel, B. Witkamp, M. Poot, H.B. Meerwaldt, L.P. Kouwenhoven, H.S.
J. van der Zant, Science \textbf{325}, 1103 (2009).

\end{thebibliography}
\end{document}